\documentclass[twocolumn,aps,prl,showpacs]{revtex4-1}
\usepackage[T1]{fontenc}
\usepackage[latin9]{inputenc}
\usepackage{amsmath}
\usepackage{amssymb}
\usepackage{graphicx}

\makeatletter

\usepackage{dcolumn}
\usepackage{bm}
\usepackage{float}

\makeatother

\begin{document}

\title{Scaling of Clusters near Discontinuous Percolation Transitions in
Hyperbolic Networks}

\author{Vijay Singh and Stefan Boettcher}

\affiliation{Dept.~of Physics, Emory University, Atlanta, GA, 30322;
  USA}

\begin{abstract}
We investigate the onset of the discontinuous percolation transition
in small-world hyperbolic networks by studying the systems-size scaling
of the typical largest cluster approaching the transition, $p\nearrow p_{c}$.
To this end, we determine the average size of the largest cluster
$\left\langle s_{{\rm max}}\right\rangle \sim N^{\Psi\left(p\right)}$
in the thermodynamic limit using real-space renormalization of cluster
generating functions for bond and site percolation in several models
of hyperbolic networks that provide exact results. We determine that
all our models conform to the recently predicted behavior regarding
the growth of the largest cluster, which found diverging, albeit sub-extensive,
clusters spanning the system with finite probability well below $p_{c}$
and at most quadratic corrections to unity in $\Psi\left(p\right)$
for $p\nearrow p_{c}$. Our study suggest a large universality in
the cluster formation on small-world hyperbolic networks and the potential
for an alternative mechanism in the cluster formation dynamics at
the onset of discontinuous percolation transitions.
\end{abstract}
\maketitle

\section{Introduction\label{sec:Introduction}}

Small-world hierarchical networks have generated much interest as
models for the prevalent hierarchical organization in complex networks
because they yield exact results for statistical models~\citep{Trusina04,Hinczewski06,SWPRL,Clauset08,Dorogovtsev08}.
These recursive structures provide deeper insights into the nonlinear
behavior caused by small-world connections, compared to some presumed
network ensemble that often requires approximate or numerical methods.
Work on percolation~\citep{PhysRevE.75.061102,Boettcher09c,Boettcher11d,Hasegawa13c,PhysRevE.82.011113},
the Ising model~\citep{Bauer05,Hinczewski06,Boettcher10c,Baek11},
and the Potts model~\citep{Khajeh07,Nogawa12} have shown that critical
behavior once thought to be exotic and model-specific~\citep{Dorogovtsev08}
can be universally described near the transition point~\citep{BoBr12,PhysRevLett.108.255703}
for a large class of hierarchical networks with hyperbolic properties.
In a hyperbolic structure, sites are typically randomly connected
but possess a hierarchical organization of sites that allows to identify
a few sites harboring many small-world bonds as central while an extensive
portion of sites with less access resides on the periphery~\citep{Hasegawa10b,PhysRevE.82.036106}.
Such structures are common in disordered materials~\citep{Wales03,Fischer08},
human organizations\citep{Trusina04}, information and communication
networks~\citep{Boguna09,PhysRevE.82.036106}, or neural networks
\citep{Meunier09,Moretti13}. However, in scale-free hyperbolic networks
\citep{Boguna11} there appears to be no threshold against the onset
of percolation.

Here, we extend the discussion of universality on such networks by
studying the emergence of the discontinuous transition recently found
in ordinary percolation~\citep{Boettcher11d}. Due to the discovery
of percolation transitions that first appeared to be ``explosive''
\citep{Achlioptas09,ISI:000286751500010,Riordan11}, the dynamics
of cluster formation at the onset of such a transition has been the
focus of much research~\citep{FriedmanLandsberg09,ISI:000291093600009,ISI:000279888400004,ISI:000282681400006,ISI:000288390400004}.
While details of the cluster size distribution $\rho(s)$ remain accessible
only to simulations, we can use the renormalization group (RG) to
determine the exact large-$N$ scaling of the average size of the
largest cluster,
\begin{equation}
\left\langle s_{{\rm max}}\right\rangle \sim N^{\Psi\left(p\right)},\label{eq:averClustSize}
\end{equation}
near the onset of the transition. Analyzing a number of different
networks for site and bond percolation, we find that the behavior
observed in Ref.~\citep{Boettcher11d} appears to be generic for hyperbolic networks. By "hyperbolic" we mean a hierarchical network with small-world properties. The hierarchy ensures the distinction between an extensive set of peripheral nodes of low centrality and ever sparser bulk nodes of increasing centrality, while small-world bonds reduce average distances to scale logarithmically with system size. In all cases, here or in related work~\citep{Hasegawa10b,Nogawa13},
it is found that within hyperbolic networks the cluster size exponent
$\Psi(p)$ defined in Eq.~\eqref{eq:averClustSize} depends on the
percolation parameter $p$ in a nontrivial manner and has only quadratic
or higher-order corrections in its approach to an extensive cluster,
$\Psi\to1$, at the transition, $p\to p_{c}$. This would suggest
the emergence of a dominant, albeit sub-extensive, cluster long before
the transition is reached.

\begin{figure*}
\includegraphics[bb=0bp 250bp 792bp 400bp,clip,scale=0.65]{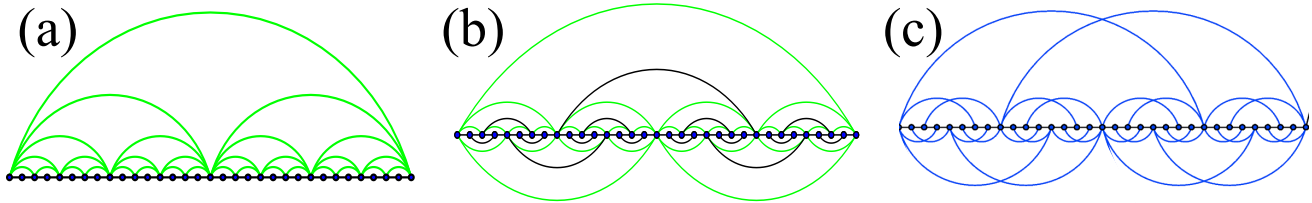}\caption{\label{fig:Networks}Depiction of hierarchical networks: (a) MK1,
(b) HN5, and (c) HNNP. For all networks the recursive pattern that
scales to the thermodynamic limit is evident. Each network features
regular geometric structures, such as a one-dimensional backbone,
and a distinct set of small-world links. While MK1 and HN5 are planar,
HNNP is non-planar.}
\end{figure*}

Such a non-linear approach towards the transition contrasts with the
behavior of the equivalent exponent, defined via the susceptibility,
on the same networks near the critical temperature for the Ising model
{[}cite future work{]}, and also with the predictions of the universal
theory for these transitions~\citep{BoBr12}, which would obtain a
linear correction generically. In a companion Communication, we will
illuminate the connection between Ising and percolation critical behavior
on these networks using the $q$-state Potts model in its analytic
continuation for non-integer values of $q$. There, we find that the
quadratic corrections persist for all $q<2$, including percolation
($q\to1$) merely as a special case. Only when $q\geq2$, including
the Ising model ($q=2$) as the $marginal$ case, do linear corrections
dominate. In the future, we will extend our Potts-model analysis to
entire families of complex networks.

This paper is organized as follows: In the following 
Sec.~\ref{sec:Small-World-Hyperbolic-Networks},
we introduce the networks used in our current study. Then, in 
Sec.~\ref{sec:ClusterRGreview}, we first review the RG-methods used to
analyze the bond percolation transition for the case previously considered
in Ref.~\citep{Boettcher11d} and then apply the same techniques in
Sec.~\ref{sec:Cluster-Size-Scaling-for-Hanoi} to the Hanoi networks;
we extract the exact quadratic corrections for bond percolation in
the cluster size exponent $\Psi$ for these networks while deferring
many of the technical details of the calculation to the Appendix.
In Sec.~\ref{sec:-Site-Perco} we show that such non-linear corrections
also characterize the site percolation transition. In 
Sec.~\ref{sec:Conclusions},
we finish with our conclusions and suggestions for future work.

\begin{figure}[b!]
\includegraphics[width=0.7\columnwidth]{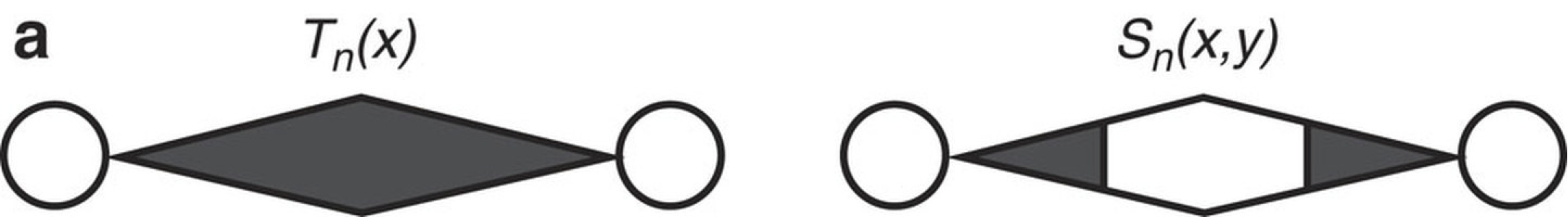} \caption{\label{MK1_defn} Diagrammatic definition of generating functions
$T_{n}(x)$ and $S_{n}(x,y)$ in Eqs.~\eqref{eq:MK1rec} for MK1 in
Fig.~\ref{fig:Networks}. End sites are represented by open circles
and clusters by shaded areas. $T_{n}(x)$ consists of one spanning
cluster, labeled $x$, which connects both end-sites and $S_{n}(x,y)$
consists of two non-spanning clusters, $x$ and $y$, each connected
to one end-site. Isolated clusters not containing either of the end
sites are ignored. }
\end{figure}

\section{Small-World Hyperbolic Networks\label{sec:Small-World-Hyperbolic-Networks}}

The models we are studying here are familiar hierarchical networks
that have become popular because they provide exact results for complex
processes by way of the real-space renormalization group. MK1, depicted
in Fig.~\ref{fig:Networks}(a), is the one-dimensional version of
the small-world Migdal-Kadanoff hierarchical diamond lattice~\citep{Hinczewski06},
which has been used previously to prove the existence of the discontinuous
transition in ordinary percolation~\citep{Boettcher11d}. MK1 is recursively
generated starting with two sites connected by a single edge at generation
$n=0$. Each new generation recursively combines two sub-networks
of the previous generation and adds single edge connecting the end
sites. As a result, the n$^{th}$ generation contains $2^{n}+1$ vertices,
$2^{n}$ backbone bonds, and $2^{n}-1$ small-world bonds.

To show that this discontinuity persists for more complicated but
hierarchical structures, we consider here also the Hanoi networks
HN5 and HNNP, also shown in Fig.~\ref{fig:Networks}(b-c). A similar
recursive procedure as described above for MK1 is also applied to
obtain each new generation, however, due to their more complicated
structure their basic building block at $n=0$ consists of a triangle
of three sites. For these Hanoi networks, the existence of a non-trivial
bond-percolation transition has been demonstrated previously~\citep{Boettcher09c}.
HN5 is similar to MK1 but requires a coupled system of RG-recursions.
It also can be easily adapted to complement previous investigations
of site-percolation~\citep{Hasegawa13b} in a non-trivial fashion.
HNNP is special in that it is a non-planar graph, and aspect that
is missing from other hierarchical networks.

\section{Review of Cluster Renormalization in Bond Percolation \label{sec:ClusterRGreview}}

Before we apply it to calculate exact expressions for the scaling
of the average cluster size for HN5 and HNNP in the next section,
we first review briefly the formalism needed to analyze the average
cluster size near the bond-percolation transition, as used for MK1
in Ref.~\citep{Boettcher11d}. While a full understanding the dynamics
of cluster formation near the discontinuous percolation transition
requires knowledge of the entire cluster-size distribution, already
the average size of the largest cluster $\left\langle s_{{\rm max}}\right\rangle _{n}$
at generation $n$ provides profound insights. In particular, we will
be focused on the system-size scaling of $\left\langle s_{{\rm max}}\right\rangle _{n}$
for $p\to p_{c}$. In the following, we derive $\left\langle s_{{\rm max}}\right\rangle _{n}$
using cluster generating functions.

\subsection{Cluster Generating Function for MK1:\label{sub:Cluster-Generating-function MK1}}

\begin{figure}
\includegraphics[width=1\columnwidth]{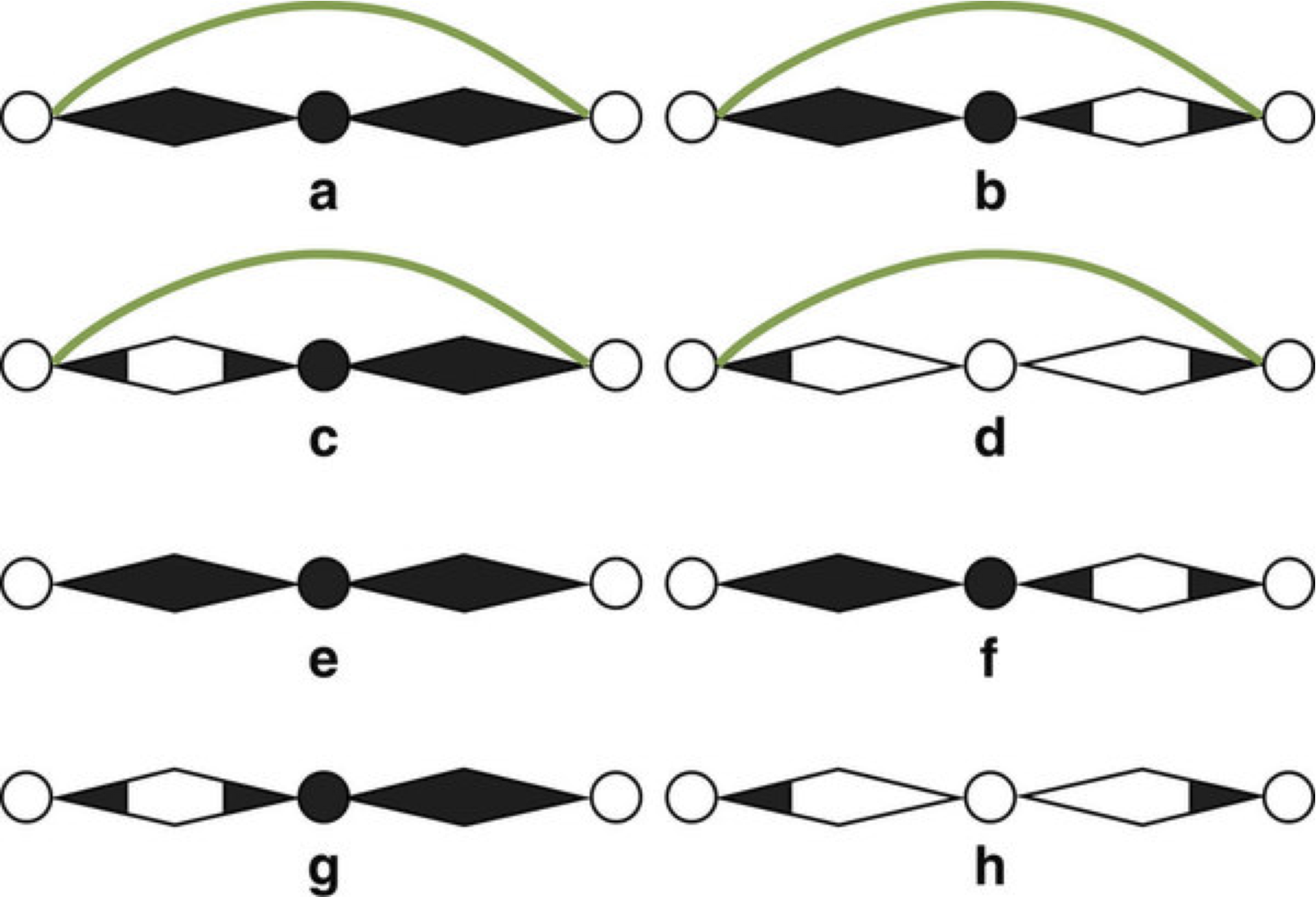} \caption{\label{MK1_3point} Diagrammatic evaluation of generating functions
for MK1. All graphlets contributing to $T_{n+1}(x)$ and $S_{n+1}(x+y)$
in the nth generation. Graphlets (\textbf{a-e}) have end-to-end connections
and contribute to $T_{n+1}(x)$ while (\textbf{f-h}) contribute to
$S_{n+1}(x,y)$. The contribution of each graphlet is (\textbf{a})
$xpT_{n}^{2}(x)$ (\textbf{b}) $xpT_{n}(x)S_{n}(x,x)$ (\textbf{c})
$xpT_{n}(x)S_{n}(x,x)$ (\textbf{d}) $pS_{n}(x,1)S_{n}(1,x)$ (\textbf{e})
$x(1-p)T_{n}^{2}(x)$ (\textbf{f}) $x(1-p)T_{n}(x)S_{n}(x,y)$ (\textbf{g})
$y(1-p)T_{n}(y)S_{n}(x,y)$ (\textbf{h}) $(1-p)S_{n}(x,1)S_{n}(1,y)$.
The recursion can be obtained by adding weights (\textbf{a-e}) for
$T_{n+1}(x)$ and (\textbf{f-h}) for $S_{n+1}(x,y)$ resulting in
Eq.~\eqref{eq:MK1rec}. See Appendix \ref{sub:Automated-Graph-Counting}
for an algorithm to automate the evaluation.}
\end{figure}

We review briefly the procedure described in Ref.~\citep{Boettcher11d}
for MK1. There, the generating functions were obtained by introducing
merely two quantities: the probability $t_{i}^{(n)}(p)$ that both
end-sites are connected to the same cluster of size $i$, and the
probability $s_{i,j}^{(n)}(p)$ that the left end-site is connected
to a cluster of size $i$ and the right end-site to a different cluster
of size $j$. The generating functions, as depicted in Fig.~\ref{MK1_defn},
are defined as
\begin{align}
T_{n}(x)= & \sum_{i=0}^{\infty}t_{i}^{(n)}(p)\, x^{i}\label{eq:GenFuncMK1}\\
S_{n}(x,y)= & \sum_{i=0}^{\infty}\sum_{j=0}^{\infty}s_{i,j}^{(n)}(p)\, x^{i}\, y^{j}.
\end{align}
The recursion relations for these generating functions can be obtained
by considering all possible configurations on three sites, as shown
in Fig.~\ref{MK1_3point}, taking into account the cluster sizes
as described in Ref.~\citep{Boettcher11d}. The graphlets on three
sites are assigned to the correct two-site graphlet in the next generation,
and the weights of all the graphlets that contribute to the same higher-generation
graphlet are added together to get the recursion relations, 
\begin{align}
T_{n+1}(x)=\, & xT_{n}^{2}(x)\nonumber \\
 & +p\left[2xT_{n}(x)S_{n}(x,x)+S_{n}(x,1)S_{n}(1,x)\right],\\
S_{n+1}(x,y)=\, & (1-p)\left[xT_{n}(x)S_{n}(x,y)+yT_{n}(y)S_{n}(x,y)\right.\nonumber \\
 & \left.\qquad+S_{n}(x,1)S_{n}(1,y)\right],\label{eq:MK1rec}
\end{align}
as indicated in Fig.~\ref{MK1_3point} and discussed in more detail
in Appendix \ref{sub:Counting-MK1-graphlets:}.

\subsection{Fixed Point Analysis for Average Cluster Size:\label{sub:Fixed-Point-Analysis}}

\begin{figure*}
\includegraphics[bb=0bp 200bp 792bp 400bp,clip,scale=0.65]{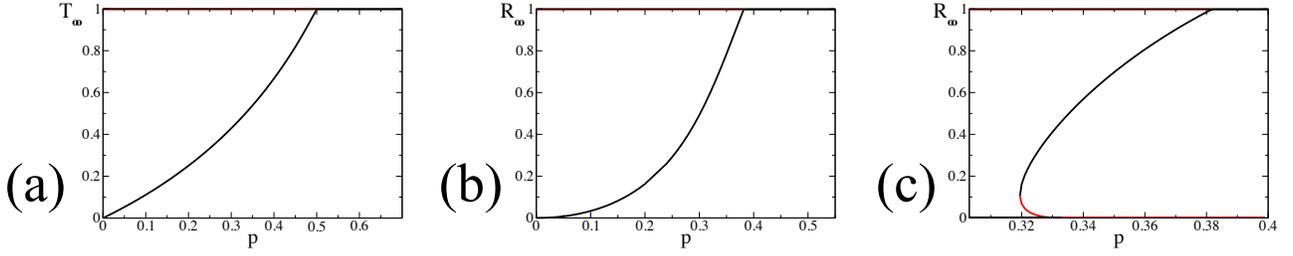}\caption{\label{fig:phasediagram}Phase diagram for the probability of a spanning
cluster (a) $T_{\infty}$ for MK1 in Eq.~\eqref{eq:MK1FP}, (b) $R_{\infty}$
for HN5 in Eq.~\eqref{eq:xyzRGflowHN5}, and (c) $R_{\infty}$ for
HNNP in Eq.~\eqref{eq:xyzRGflowHNNP} (for $x=1$), all as a function
of bond probability $p$. Black lines mark stable fixed points, and
red-shaded lines are unstable fixed point solutions. The critical
transition, at which the probability of any site to belong to the
largest cluster becomes finite and that cluster becomes extensive,
occurs exactly when the probability of a spanning cluster becomes
unity, at $p_{c}=\frac{1}{2}$ for MK1 and $p_{c}=2-\phi=0.38197\ldots$
for both, HN5 and HNNP~\citep{Boettcher09c}. However, in all cases,
there is a non-zero probability for a spanning cluster, albeit sub-extensive,
even below $p_{c}$, due to the hyperbolic nature of these hierarchical
networks. For MK1 and HN5, such a cluster can exist for all $0<p<p_{c}$,
while for HNNP it disappears below the branch-point singularity at
$p_{l}=0.31945\ldots$. Note that in each case the transition occurs
at the intersection of two lines of \emph{stable} fixed points.}
\end{figure*}

The recursion equations in Eq.~\eqref{eq:MK1rec} can be simplified
by combination them into a vector $\vec{V}_{n}\left(x\right)=[T_{n}\left(x\right),S_{n}\left(x,x\right),S_{n}\left(x,1\right)]$
of distinct observables, where we focus on the largest cluster $x$
only. The RG can now be written as
\begin{align}
\vec{V}_{n+1}(x)=\vec{F}\left(\vec{V_{n}}(x),x\right)\label{MK1_vec_genfun}
\end{align}
for the nonlinear vector-function $\vec{F}$ that derives from Eqs.
\ref{eq:MK1rec}. As Eq.~\eqref{eq:GenFuncMK1} suggest, the average
size of a spanning cluster (which dominate in the cluster-size distribution)
is generated by $\left\langle s\right\rangle \sim T_{n}^{\prime}\left(x=1\right)$;
any form of $S_{n}$ does not affect to the spanning cluster and its
contributions prove subdominant. We obtain $T_{n}^{\prime}\left(x=1\right)$
in terms of $T_{n}=T_{n}\left(x=1\right)$ and $p$ by linearizing
the recursion relation in Eq.~\ref{MK1_vec_genfun} 
\begin{align}
\frac{\partial\vec{V}_{n+1}}{\partial x}=\frac{\partial\vec{F}}{\partial\vec{V}}\left(\vec{V_{n}}\right)\cdot\frac{\partial\vec{V}_{n}}{\partial x}+\frac{\partial\vec{F}}{\partial x}\left(\vec{V_{n}}\right),\label{MK1_first_order}
\end{align}
near $x=1$. Eq.~\eqref{MK1_vec_genfun} itself at $x=1$ (where $S_{n}=1-T_{n}$)
reduces for MK1 in each component of $\vec{V}$ to 
\begin{align}
T_{n+1}=p+\left(1-p\right)T_{n}^{2}\qquad\left(T_{0}=p\right)
\end{align}
with fixed point $T_{\infty}=\lim_{n\to\infty}T_{n}$
\begin{align}
T_{\infty}\left(p\right)=\left\{ \begin{matrix}\frac{p}{(1-p)} &  & 0\le p<\frac{1}{2}\\
1 &  & \frac{1}{2}\le p\le1,
\end{matrix}\right.\label{eq:MK1FP}
\end{align}
providing the critical point $p_{c}=\frac{1}{2}$, where any spanning
cluster also becomes extensive, see Fig.~\ref{fig:phasediagram}(a). 

Ignoring the subdominant inhomogeneity in Eq.~\eqref{MK1_first_order},
the remaining homogeneous linear system gives the dominant contribution
for $V_{\infty}^{\prime}$, i.e. $T_{\infty}^{\prime},S_{\infty}^{\prime}$.
The largest eigenvalue $\lambda$ of the coefficient-matrix $\frac{\partial\vec{F}}{\partial\vec{V}}\left(\vec{V_{\infty}}\right)$
at the fixed point $T_{\infty}\left(p\right)$ becomes for MK1
\begin{align}
\lambda=\left\{ \begin{matrix}\frac{1+3p-4p^{2}}{2(1-p)}+\sqrt{\frac{1-p(1-4p)^{2}}{4(1-p)}} &  & 0\le p<\frac{1}{2}\\
2 &  & \frac{1}{2}\le p\le1.
\end{matrix}\right.\label{Mk1_det}
\end{align}
Finally, we obtain the order parameter $P_{\infty}$ as
\begin{align}
P_{\infty}=\frac{\left\langle s_{{\rm max}}\right\rangle }{N}\sim\frac{T_{\infty}^{\prime}}{N}\sim N^{\Psi\left(p\right)-1}\label{eq:OrderP}
\end{align}
with the fractal exponent\eqref{eq:PsiLambda}
\begin{equation}
\Psi\left(p\right)=\log_{2}\lambda.\label{eq:PsiLambda}
\end{equation}
Note that this implies that the largest cluster below the transition
is already diverging with a non-zero power of the system size, although
in a sub-extensive manner, $\Psi<1$ for $p<p_{c}$, such that $P_{\infty}\to0$
for $N\to\infty$. These spanning, sub-extensive clusters exist, albeit
with finite probability given by $T_{\infty}\left(p\right)$ in 
Eq.~\eqref{eq:MK1FP}, for all $0<p<p_{c}$. This behavior for hyperbolic
systems contrasts with that of regular lattices, where such sub-extensive
clusters with fractal scaling only exist for $p=p_{c}$ and $\Psi(p)\equiv0$
for $p<p_{c}$ such that all clusters remain finite or at most diverge
logarithmically in $N$.

\begin{figure*}
\includegraphics[bb=0bp 200bp 792bp 400bp,clip,scale=0.65]{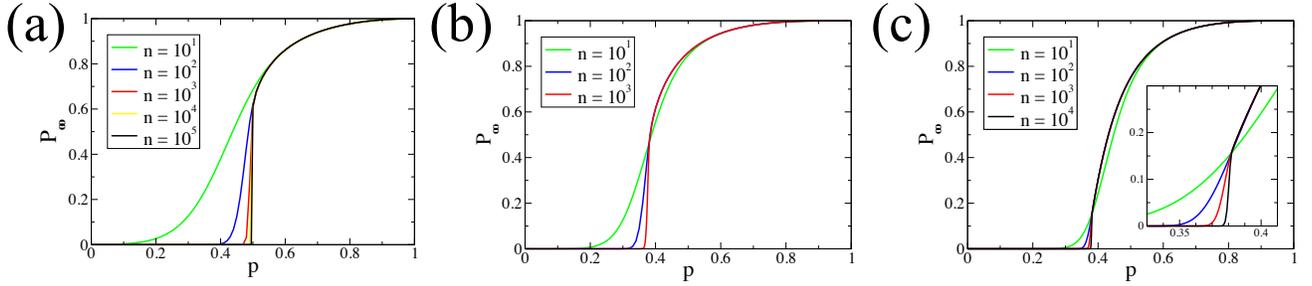}\caption{\label{fig:Pinf}Discontinuity in the percolation order parameter
$P_{\infty}(p)$ for (a) MK1, (b) HN5, and (c) HNNP, each for $n=10^{k}$
iterations for some integer $k$. In each case, $P_{\infty}$ converges
slowly to zero just below $p_{c}$, and at $p_{c}$, $P_{\infty}$
changes discontinuously. The discontinuity decreases left to right,
and is barely visible for HNNP, see inset. }
\end{figure*}

In Fig.~\ref{fig:Pinf}(a), we show a plot of $P_{\infty}(p)$ for
MK1 evaluated after $n=10^{k}$ iterations using Eq.~(\ref{MK1_first_order})
displayed for $k=1,...,5$ corresponding to system sizes up to $N\simeq2^{n}\sim10^{3010}$
sites. $P_{\infty}$ converges slowly to zero for $p<p_{c}=\frac{1}{2}$.
At and above $p_{c}$, it can be shown using Eq.~\ref{MK1_first_order}
that $T_{n}^{\prime}$ is monotonically increasing with $n$ while
being bounded above by $1$, thus the order parameter is positive
definite for $\frac{1}{2}\le p<1$. The order parameter $P_{\infty}$
changes discontinuously from $0$ to $0.609793...$ at $p=p_{c}$
and converges to $1$ for $p\rightarrow1$. A more detailed discussion,
including a proof of the discontinuity, is provided in Ref.~\citep{Boettcher11d}.

\subsection{Scaling Behavior near the Transition\label{sub:Average-Cluster-Size}}

From Eqs.~(\ref{Mk1_det}-\ref{eq:PsiLambda}) it is now easy to determine
the scaling behavior for the average cluster size near the transition.
By expanding the eigenvalue $\lambda$ in Eq.~\eqref{Mk1_det} for
$p\to p_{c}$ from below, we find that the leading behavior only has
quadratic corrections, and inserting into Eq.~\eqref{eq:PsiLambda}
results in
\begin{equation}
\Psi(p)\sim1-\frac{8}{{\rm \ln2}}\left(p-p_{c}\right)^{2},\qquad p\nearrow p_{c}=\frac{1}{2},
\end{equation}
which rapidly approaches unity. This implies that the largest (spanning)
cluster that dominates the distribution is nearly extensive already
much before the discontinuous transition is reached. RG can only determine
the probability $T_{\infty}$ and average size $\left\langle s_{{\rm max}}\right\rangle \sim T_{\infty}^{\prime}$
of the spanning cluster. Their sub-extensive nature for $p<p_{c}$
would allow in principle for a diverging number of such clusters.
Our simulations show that already for small systems the largest cluster
is almost certainly connected to at least one end-site near $p_{c}$.
(In fact, for MK1 we could have just as well defined $\left\langle s_{{\rm max}}\right\rangle \sim T_{\infty}^{\prime}+{\cal S}_{\infty}^{\prime}$
to account not just for spanning but all end-site connected clusters,
without affecting the scaling.) However, as we will see for HNNP,
the non-extensive clusters further below $p_{c}$ may well be purely
internal, with zero probability of spanning between any end-sites.

\begin{figure*}
\includegraphics[bb=0bp 200bp 792bp 400bp,clip,scale=0.65]{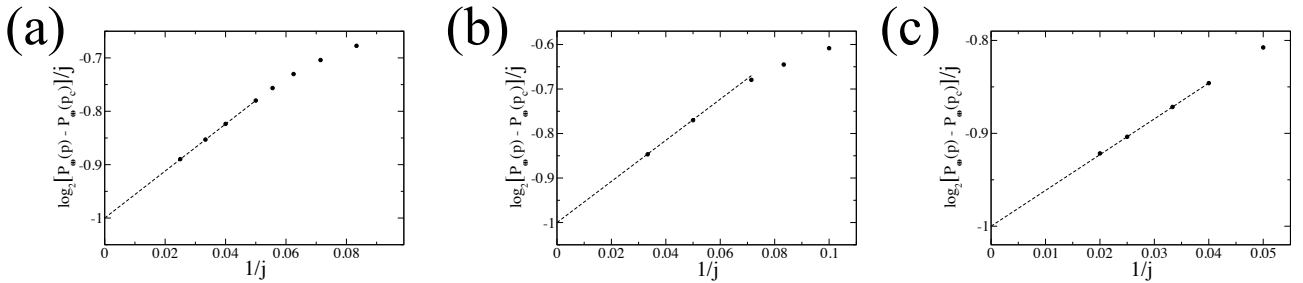}\caption{\label{fig:beta} Scaling of the order parameter $P_{\infty}(p)$
for $p\searrow p_{c}$ according to Eq.~\eqref{eq:BetaScaling} for
(a) MK1, (b) HN5, and (c) HNNP. In each case, taking $p-p_{c}=\frac{1}{2^{j}}$,
we plot
$\log_2\left[P_{\infty}\left(p\right)-P_{\infty}\left(p_{c}\right)\right]/j$
vs.~$1/j$ which linearly extrapolates to $\beta\sim1$ as the intercept
at $j\to\infty$, i.e., $p\to p_{c}$ . }
\end{figure*}

In light of the discussion regarding universal behavior in hyperbolic
networks~\citep{BoBr12,Nogawa13}, it is interesting to also explore
the scaling behavior of the order parameter on its approach to the
discontinuity from above the transition. Numerically, with the RG,
we find that a fit to 
\begin{equation}
P_{\infty}\left(p\right)\sim P_{\infty}\left(p_{c}\right)+A\left(p-p_{c}\right)^{\beta}\qquad\left(p\searrow p_{c}\right)\label{eq:BetaScaling}
\end{equation}
is quite consistent with a simple, linear approach, i.e., $\beta=1$,
see Fig.~\ref{fig:beta}(a).

\section{Cluster-Size Scaling for Hanoi Networks\label{sec:Cluster-Size-Scaling-for-Hanoi}}

In the following, we will apply the formalism from Sec.~\ref{sec:ClusterRGreview}
to the Hanoi networks HN5 and HNNP in Fig.~\ref{fig:Networks}(b-c).
Their phase diagram, as shown in Fig.~\ref{fig:phasediagram}(b-c),
has already been discussed in Ref.~\citep{Boettcher09c}. To obtain
their average cluster size requires the automated algorithm developed
in the Appendix, due to the substantial combinatorial effort to enumerate
their conformations. We will focus here on the more interesting case
of HNNP first and then merely report equivalent results for HN5, without
the details. 

Despite of the added complexity, we find remarkably similar results
near the transition for these networks, as compared to MK1, and only
some distinctly interesting features for HNNP in the ``patchy''
regime below $p_{c}$. Such robust behavior suggests universal features
\citep{BoBr12,Nogawa13}, which can be traced back to the fundamental
phase diagram shared by all three networks, as is evident from Fig.
\ref{fig:phasediagram}. For comparison, this bond-percolation behavior
is not shared by another hierarchical network, MK2, which mutatis
mutandis has quite a distinct phase diagram~\citep{Boettcher09c,Berker09},
leading instead to a BKT transition. See Ref.~\citep{Nogawa13} for
an interpolation between both cases.

In the Appendix, Sec.~\ref{sub:Cluster-Generating-Function HNNP},
we show how to obtain the RG-recursions for the cluster generating
functions. While otherwise similar to the discussion in Sec.~\ref{sub:Cluster-Generating-function MK1},
HNNP (as well as HN5) requires four such functions to account for
all possibilities, of having clusters linking any combination of three
end-sites or remain isolated, even after accounting for all symmetries
of the network. The resulting recursions, Eqs.~\eqref{eq:xyzRGflowHNNP},
are similar to those for MK1 in Eqs.~\eqref{eq:MK1rec}, although
rather more involved. In the end, we only care for the dominant cluster,
which we label $x$, and consider each possible contribution from
one RG-step to the next while disregarding sub-dominant clusters by
setting $y=z=1$. Note that even clusters that are disconnected from
any end-site at one step could significantly contribute at the next
via the small-world bonds that are linking graphlets between consecutive
RG-steps. In the end, we can identify \emph{ten} distinct observables
that form a closed set of recursions. When combined into a single
vector,
\begin{eqnarray}
\vec{V}_{n}(x) & = & \left[R_{n}(x),S_{n}(x,x),S_{n}(x,1),U_{n}(x,x),\right.\nonumber \\
 &  & \quad U_{n}(x,1),N_{n}(x,x,x),N_{n}(x,x,1),\label{eq:VnHNNP}\\
 &  & \left.\, N_{n}(x,1,x),N_{n}(x,1,1),N_{n}(1,x,1)\right],\nonumber
\end{eqnarray}
these satisfy the equivalent recursion in~\eqref{MK1_vec_genfun},
with the nonlinear RG-flow given by Eqs.~\eqref{eq:xyzRGflowHNNP}.

To zeroth order, at $x=1$, Eq.~\eqref{MK1_vec_genfun} gives the
recursion relation for percolation of the HNNP graph as derived in
Ref.~\citep{Boettcher09c}. The coupled recursion relations in ($R_{n},S_{n},U_{n},N_{n}$)
result in the roots of a sextic polynomial, which can be solved numerically
to get the probability of, say, the spanning cluster $R_{\infty}$
between the end-sites. Fig.~\ref{fig:phasediagram}(c) gives the
phase diagram for HNNP representing the solutions of the sextic equation,
which correspond to the probability $R_{\infty}$ for $0<p<1$. HNNP
provides a unique example of a network in which the probability of
the dominant cluster to touch \emph{any} end-site vanish below some
finite value $0<p_{l}<p_{c}$. In Ref.~\citep{Boettcher09c} this
was interpreted as a second, lower, critical point, where below $p_{l}$
neither a spanning nor an extensive cluster exists while between $p_{l}$
and $p_{c}$ at least a spanning cluster exists that does not need
to be extensive, due to the hyperbolic structure of the network. That
spanning cluster becomes extensive only above $p_{c}$, the true critical
percolation point with non-zero order parameter, $P_{\infty}>0$.
However, as was shown in Ref.~\citep{Hasegawa13c}, even below the
non-zero $p_{l}$ in HNNP a diverging cluster remains and $\Psi(p)$
defined in Eq.~(\ref{eq:averClustSize}) remains positive for all
$p>0$. At $p_{l}$, $\Psi(p)$ merely jumps discontinuously to a
lower but finite value, yet, diverging clusters that connect end-sites
are almost certainly absent. Any diverging cluster is fully contained
inside HNNP. 

\begin{figure}
\includegraphics[bb=0bp 0bp 792bp 612bp,clip,width=1\columnwidth]{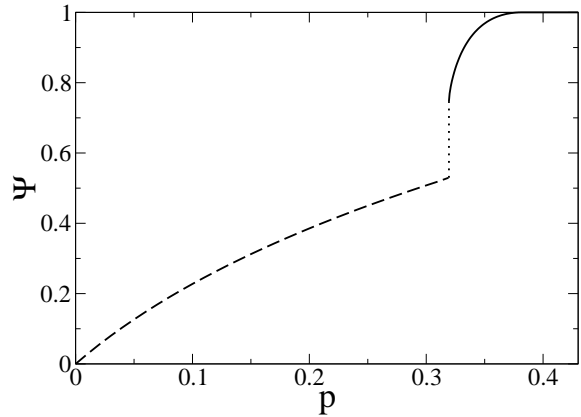}\caption{Plot of the fractal exponent $\Psi(p)$ for HNNP. The behavior of
$\Psi(p)$ for $p_{l}<p<p_{c}=0.38197\ldots$ (full line) is obtained
by exact evaluation of the Jacobian matrix, which develops a branch-point
singularity at $p_{l}=0.31945\ldots$. Ref.~\citep{Hasegawa13c} has
provided a lower bound, \textbf{$\Psi(p)=\log_{2}\left(1+\sqrt{1+8p}\right)-1$
}for $p<p_{l}$ (dashed line), suggesting a discontinuity in the scaling
of the largest cluster at $p_{l}$ (dotted line) when spanning clusters
emerge. \label{Psi_HNNP} }
\end{figure}

The nature of the largest cluster can be studied by looking at the
first-order term in the Taylor expansion, Eq.~\ref{MK1_first_order},
of the vector $\vec{V}_{n}(x)$ in Eq.~\ref{eq:VnHNNP}. For HNNP
the Jacobian $\frac{\partial\vec{F}}{\partial\vec{V}}\left(\vec{V}_{n}\right)$
at $x=1$ consists now of a $10\times10$ matrix and the inhomogeneity
is a $10\times1$ matrix. For large system sizes ($n\to\infty$) at
$x=1$, it can be shown that the inhomogeneity is subdominant, leaving
a homogeneous equations. As before, the largest eigenvalue of the
Jacobian gives the scaling exponent $\Psi(p)$ for the largest cluster
in the network from Eq.~\eqref{eq:PsiLambda}, as shown in Fig.~\ref{Psi_HNNP}.
It shows that $\Psi(p)<1$ for $p_{l}<p<p_{c}$, but $\Psi(p)$ drops
to zero discontinuously at $p_{l}$ and vanishes for $p<p_{l}=0.31945\ldots$,
since the cluster measured by the RG is conditioned on being rooted
at an end-site. The RG misses diverging clusters that that do not
span the network which apparently dominate below $p_{l}$~\citep{Nogawa13}.
In any case, since $\Psi(p)<1$, Eq.~\eqref{eq:OrderP} ensures that
$P_{\infty}\equiv0$ for all $0\leq p<p_{c}$.

Near $p_{c}=2-\phi$, where $\phi=\left(\sqrt{5}+1\right)/2$ is the
``golden section'', we again find a percolation transition with
a discontinuous jump in the order parameter $P_{\infty}$. By evolving
the recursion equations \eqref{MK1_first_order} for $V_{n}^{\prime}$
, the order parameter can be rigorously shown to have monotone convergence
to non-zero values \emph{at} and above $p_{c}$, see Fig.~\ref{fig:Pinf}(c).
For $p\nearrow p_{c}$ , the way $\Psi(p)$ approaches unity can be
found through considering the secular equation
\begin{equation}
0=\det\left\{ V_{\infty}^{\prime}-\left(2-a_{1}\epsilon+a_{2}\epsilon^{2}+\ldots\right)\times\mathbf{I}\right\} ,\label{eq:seceq}
\end{equation}
expanded in terms of $\epsilon=p_{c}-p\ll1$, where $\mathbf{I}$
is the identity matrix. Note that at $p_{c}$, the largest eigenvalue
of $V_{\infty}^{\prime}$ is $\lambda=2$, around which we expand.
Since the percolation probabilities at $p_{c}$ are given by $R_{\infty}=1,S_{\infty}=U_{\infty}=N_{\infty}=0$,
we assume an expansion of the percolation probabilities as $R_{\infty}=1-\rho_{1}\epsilon+\rho_{2}\epsilon^{2}$,
$S_{\infty}=\sigma_{1}\epsilon+\sigma_{2}\epsilon^{2}$, $U_{\infty}=\nu_{1}\epsilon+\nu_{2}\epsilon^{2}$,
and $N_{\infty}=\eta_{1}\epsilon+\eta_{2}\epsilon^{2}$. To satisfy
Eq.~\eqref{eq:seceq}, each coefficient in powers of $\epsilon$ should
be zero. As a result, we find that linear corrections to the eigenvalue
$\lambda$ vanish, i.e., $a_{1}=0$. Using conservation of probability,
$\rho_{i}+\sigma_{i}+\upsilon_{i}+\eta_{i}=0$, for each $i\geq1$
at $p=p_{c}$, we find a non-vanishing quadratic correction, $a_{2}=a_{2}(\rho_{1},\sigma_{1},\nu_{1},\eta_{1})=-\frac{5}{16}\left(38+17\sqrt{5}\right)$,
for which the second-order corrections in the percolation probabilities
proved irrelevant. Hence, Eq.~\eqref{eq:PsiLambda} yields
\begin{align}
\Psi_{{\rm HNNP}}(p)\sim1-\frac{5\left(38+17\sqrt{5}\right)}{32\log_e(2)}\left(p_{c}-p\right)^{2}+\ldots, & \quad p\nearrow p_{c}.\label{eq:PsiHNNP}
\end{align}

For HN5, by using the same cluster generating functions as for HNNP
in the Appendix, we obtain their RG recursions in \eqref{eq:xyzRGflowHN5}.
Again, the resulting equations for the cluster size are too complicated
to express or solve in closed form. But it is easy to evaluate their
phase diagram in Fig.~\eqref{fig:phasediagram}(b) for $R_{\infty}$,
as well as the order parameter $P_{\infty}$ in Fig.~\eqref{fig:Pinf}(b)
to any desired accuracy. Here, the same local analysis near $p_{c}$
as for HNNP yields for HN5:
\begin{align}
\Psi_{{\rm HN5}}(p)\sim1-\frac{5\left(677+304\sqrt{5}\right)}{484\log_e(2)}\left(p_{c}-p\right)^{2}+\ldots, & \quad p\nearrow p_{c}.\label{eq:PsiHN5}
\end{align}
As for MK1 and HNNP, almost extensive clusters in HN5 emerge well
before the transition, with $\Psi(p)$ varying quadratically. It suggests
that the quadratic dependence below $p_{c}$ might be universal for
hierarchical networks with discontinuous percolation transitions.
Above $p_{c}$, the scaling of $P_{\infty}$ in Eq.~\eqref{eq:BetaScaling}
for both, HN5 and HNNP, also provides $\beta\sim1$, as shown in Fig.~\ref{fig:beta}(b-c).

\section{Cluster Size for Site Percolation\label{sec:-Site-Perco}}

We supplement these findings with a unique result of even higher-order
behavior in the site-percolation transition of HN5 in Fig.~\ref{fig:Networks}.
The fragility of complex networks under random site-removal has recently
been studied on hierarchical networks~\citep{Hasegawa13b}. It was
shown that there is no threshold at which the network preserves an
extensive cluster, i.e., $p_{c}=1$, yet, similar quadratic corrections
in scaling to the formation of an extensive cluster for $p\to1$ are
also found there. Hence, we would expect that cluster formation near
this discontinuity is generic for both, bond- and site-percolation.
In light of this, the \emph{cubic} corrections we report here for
HN5 may provide an alternative, special case and a new clue in understanding
cluster formation. 

\begin{figure}
\includegraphics[width=0.32\columnwidth]{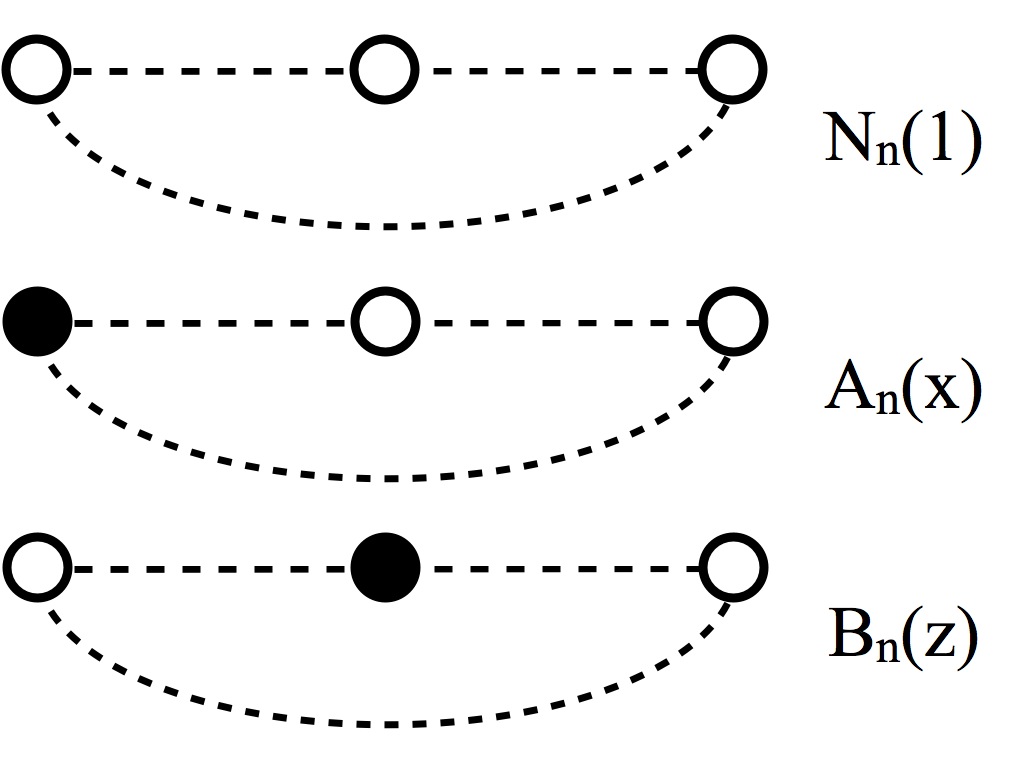}\includegraphics[width=0.32\columnwidth]{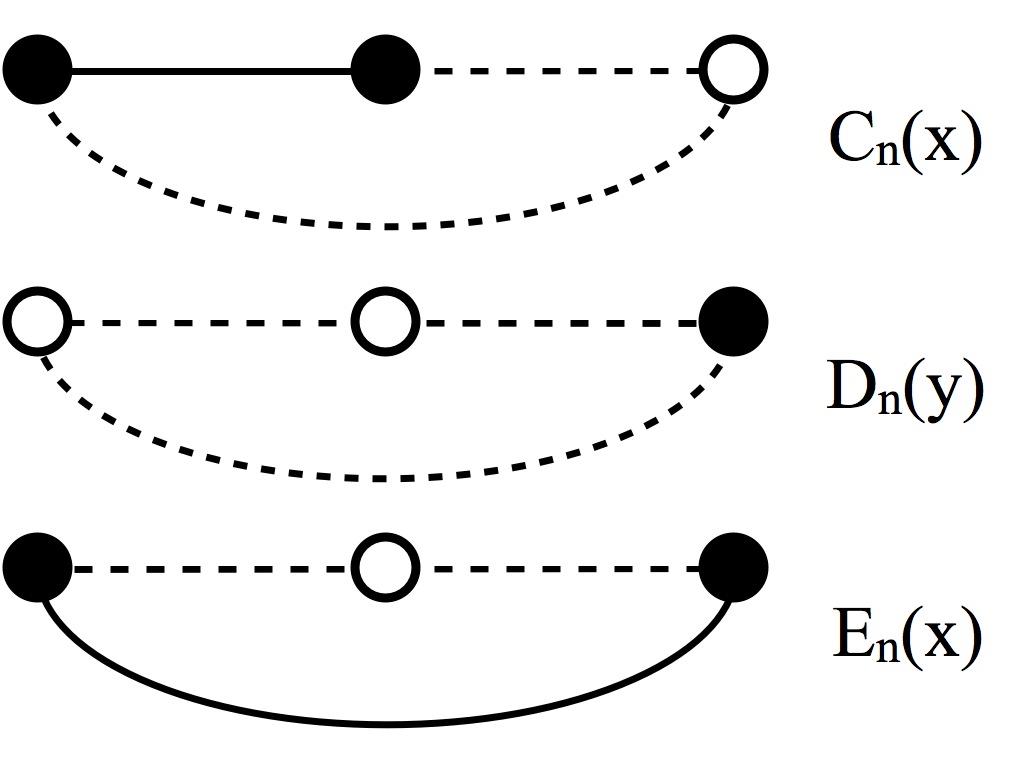}\includegraphics[width=0.32\columnwidth]{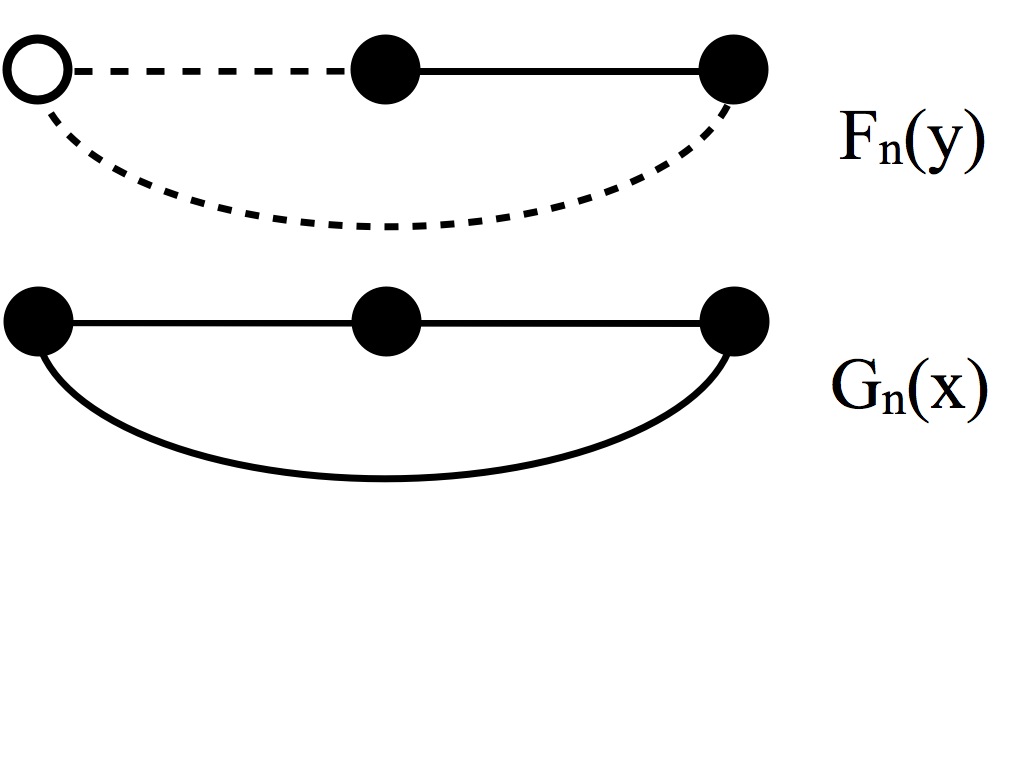}\caption{Depiction of elementary HN5 graphlets for site percolation. Listed
are all ($2^{3}=8$) three-site graphlets used in the recursive composition
of Hanoi networks. Filled (or unfilled) circles mark occupied (or
unoccupied) sites, each with independent probability $p$ (or $1-p$).
The arguments $x$, $y$, and $z$ indicate that each triangle harbors
a single cluster, represented by a polynomial generating function
in that variable. A full line corresponds to an existing connection
between occupied sites, and a dashed line is a possible, but unrealized,
connection when one adjacent site is unoccupied. Note that $A_{n}$
and $D_{n}$, and $C_{n}$ and $F_{n}$ are simply mirror images of
each other that satisfy the same recursions; hence, we can eliminate
$D_{n}$ and $F_{n}$ from the recursions in the end.}
\label{HN5_graphlets} 
\end{figure}

\begin{figure}
\includegraphics[scale=0.2]{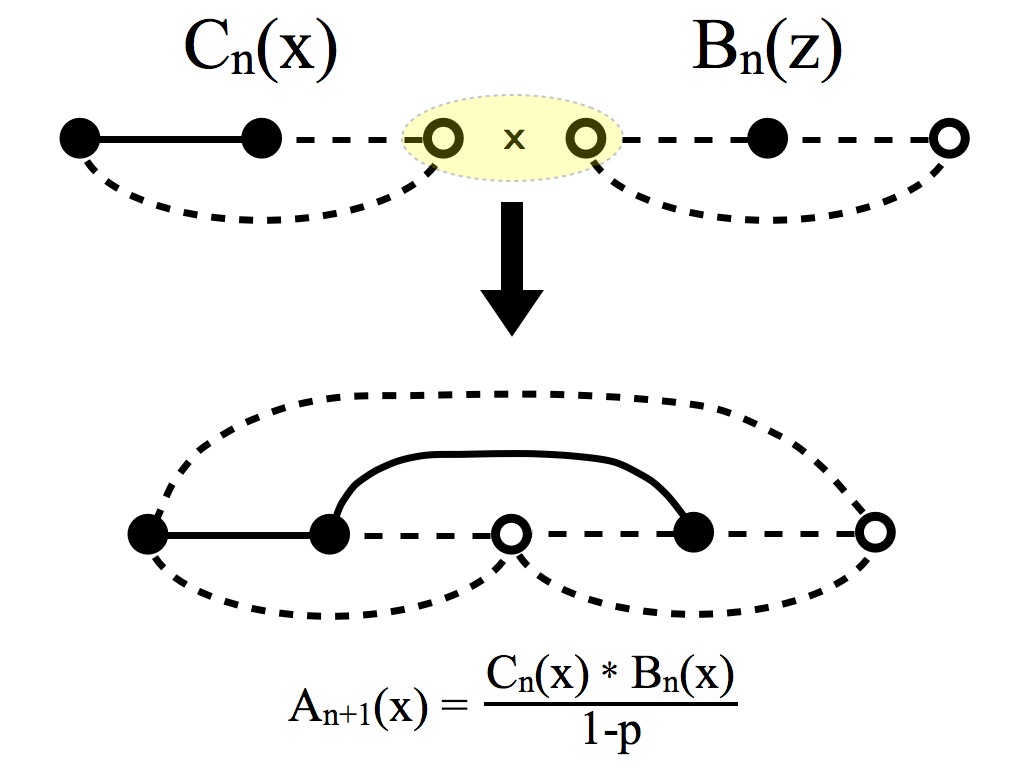}\caption{\label{fig:Demonstration-of-merging}Demonstration of the merging
of elementary graphlets into a graphlet of the next generation in
HN5, the generic five-site structure being exhibited by the lower
diagram. Here, graphlets for $C_{n}\left(x\right)$ and $B_{n}\left(y\right)$
(defined in Fig.~\ref{HN5_graphlets}) are merged by overlapping at
the highlighted inner sites that become one. Adding the new long-range
bonds, a graphlet of HN5 is formed (below). The lower one of those
bonds unifies the occupied sites left and right into a single cluster,
reducing the labeling from $x$ and $z$ into a single label $x$.
Renormalization now consists of eliminating the 2nd and 4th site and
attributing their properties to the respective root-sites (left, right
and center sites). Here, for instance, there is merely one cluster
labeled $x$ that only connects to the left root, the center and right
root remain empty. Thus, this graphlet renormalizes into the type
$A_{n+1}\left(x\right)$, also defined in Fig.~\ref{HN5_graphlets}.
The entire RG consists of evaluating such a merger for all $2^{5}=32$
possible site-occupancies in the HN5 graphlet to obtain the recursions
in Eq.~(\ref{eq:HN5_site_recursion_simple}). Of course, mergers can
only be realized when the overlapping inner sites are in the same
state; that merger has to be corrected for by dividing out $1/\left(1-p\right)$
when an empty site is over-counted, as in this example, or by $1/\left(xp\right)$
when an overlapping occupied site is over-counted. Incidentally,
this case (and its mirror image) is the \emph{only} graphlets among
all 32 for which the lower long-range bond -- the distinguishing feature
between MK1 and HN5 -- makes a difference; otherwise the site from
$B_{n}\left(y\right)$ on the right would be disconnected from any
root and would remain uncounted.}
\label{HN5_merger} 
\end{figure}

With the framework for studying bond percolation on hierarchical networks
established in Sec. \ref{sec:ClusterRGreview}, we apply the same
protocols to study site percolation. HN5 can be assembled recursively
by combining all possible triangle permutations listed in Fig.~\ref{HN5_graphlets}
through mergers as explained in Fig.~\ref{HN5_merger}. Clusters
are labeled $x$ if they at least touch the left-most root site, $y$
if they do not touch the left root but at least the right-most root
site, and $z$ if they only reach the central root site. If all root
sites are unoccupied, there are no countable clusters to label, and
the argument becomes unity. Extra small-world bonds, as in the construction
of HN5 in Fig.~\ref{HN5_merger}, may combine clusters, which entails
a relabeling dictated by the same priority.

Based on these rules explained in Fig.~\ref{fig:Demonstration-of-merging},
applied to the merger of all possible graphlets in Fig.~\ref{HN5_graphlets},
the following RG-recursions for the cluster generating functions are
derived:
\begin{eqnarray}
N_{n+1}\left(1\right) & = & \frac{1}{1-p}\left[N_{n}\left(1\right)+B_{n}\left(1\right)\right]^{2},\label{eq:HN5_site_recursion_simple}\\
A_{n+1}\left(x\right) & = & \frac{1}{1-p}\left\{ \right.\left[A_{n}\left(x\right)+C_{n}\left(x\right)\right]\left[N_{n}\left(1\right)+B_{n}\left(1\right)\right]\nonumber \\
 &  & +C_{n}\left(x\right)\left[B_{n}\left(x\right)-B_{n}\left(1\right)\right]\left.\right\} ,\nonumber \\
B_{n+1}\left(z\right) & = & \frac{1}{xp}\left[A_{n}\left(z\right)+C_{n}\left(z\right)\right]^{2},\nonumber \\
C_{n+1}\left(x\right) & = & \frac{1}{xp}\left[A_{n}\left(x\right)+C_{n}\left(x\right)\right]\left[E_{n}\left(x\right)+G_{n}\left(x\right)\right],\nonumber \\
E_{n+1}\left(x\right) & = & \frac{1}{1-p}\left[A_{n}\left(x\right)+C_{n}\left(x\right)\right]^{2},\nonumber \\
G_{n+1}\left(x\right) & = & \frac{1}{xp}\left[E_{n}\left(x\right)+G_{n}\left(x\right)\right]^{2}.\nonumber 
\end{eqnarray}
Here, we already have exploited a mirror symmetry between $A_{n}$
and $D_{n}$ and between $C_{n}$ and $F_{n}$ to simplify the equations.
The initial conditions for these RG-recursions are:
\begin{eqnarray}
N_{0}\left(x\right)  = \left(1-p\right)^{3},~~~~\qquad&&A_{0}\left(x\right)  =  xp\left(1-p\right)^{2},\nonumber \\
B_{0}\left(z\right)  =  zp\left(1-p\right)^{2},~\qquad&&C_{0}\left(x\right)  =  x^{2}p^{2}\left(1-p\right),\nonumber  \\
E_{0}\left(x\right)  =  x^{2}p^{2}\left(1-p\right),\qquad&&G_{0}\left(x\right)  =  x^{3}p^{3}.\label{eq:IC}
\end{eqnarray}
Unlike the recursions for the bond-cluster generating functions, for
example, Eq.~\eqref{eq:MK1FP} for MK1, here the \emph{site}-cluster
generating functions themselves do not satisfy interesting recursions
at $x=1$. For instance, $A_{n}\left(1\right)=A_{0}\left(1\right)=p\left(1-p\right)$
for all $n$ merely reflects the \emph{defining} feature of the site-percolation
cluster $A_{n}\left(x\right)$ of being occupying the left end-site
but not the right end-site.

Note that \emph{without} the seemingly minor distinction between $B_{n}\left(x\right)$
and $B_{n}\left(1\right)$ in the $A_{n+1}$-relation, as explained
in Fig.~\ref{fig:Demonstration-of-merging}, we could drastically
reduce the recursions further by defining 
\begin{eqnarray}
T_{n}\left(x\right) & = & \frac{1}{x^{2}p^{2}}\left[E_{n}\left(x\right)+G_{n}\left(x\right)\right],\\
S_{n}\left(x\right) & = & \frac{1}{xp\left(1-p\right)}\left[A_{n}\left(x\right)+C_{n}\left(x\right)\right],\nonumber 
\end{eqnarray}
which converts Eqs.~(\ref{eq:HN5_site_recursion_simple}) into those
for MK1 in Ref.~\citep{Hasegawa13b}. Instead, we have to evolve the
entire set of five $x$-dependent relations for the RG-flow in Eqs.
(\ref{eq:HN5_site_recursion_simple}).

\begin{figure}
\includegraphics[width=1\columnwidth]{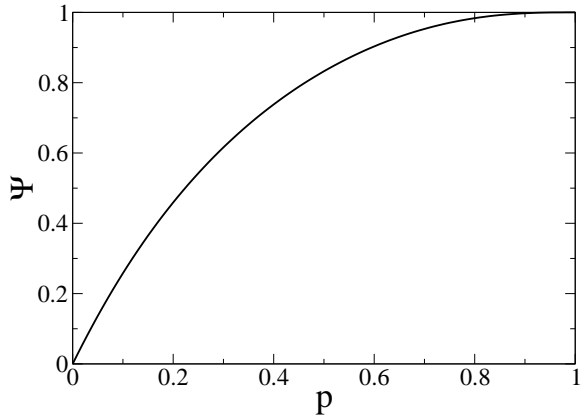}\caption{\label{fig:Plot-of-PsiHN5}Plot of $\Psi=\log_{2}\lambda$ in HN5
as a function of the site-occupation probability $p$, obtained from
largest solution of the eigenvalue Eq.~(\ref{eq:lambdaHN5}). Noticeable
is the slow rise for $p\to1^{-}$ derived in Eq.~(\ref{eq:Psi1Hn5}). }
\end{figure}

Defining
\begin{eqnarray}
\vec{V}_{n}(x) & = & \left[A_{n}(x),B_{n}(x),C_{n}(x),E_{n}(x),G_{n}(x)\right]\label{eq:VnHN5site}
\end{eqnarray}
 and following the discussion in Sec.~\ref{sec:ClusterRGreview},
we obtain from Eqs.~\eqref{eq:HN5_site_recursion_simple} at $x=1$:
\begin{equation}
\frac{\partial\vec{F}}{\partial\vec{V}}\left(\vec{V}_{\infty}\right)=\left(\begin{array}{ccccc}
1-p & p^{2} & 1-p & 0 & 0\\
2\left(1-p\right) & 0 & 2\left(1-p\right) & 0 & 0\\
p & 0 & p & 1-p & 1-p\\
2p & 0 & 2p & 0 & 0\\
0 & 2\left(1-p\right) & 0 & 0 & 0
\end{array}\right).\label{eq:dFdV_HN5site}
\end{equation}
where we used the IC in Eqs.~(\ref{eq:IC}) and the fact explained
above that $\vec{V}_{n}\left(1\right)=\vec{V}_{0}\left(1\right)$
for any $n$ for site-percolation generating functions. Then, the
largest eigenvalue is the largest root of the cubic equation
\begin{equation}
0=4p^{3}-4p^{4}+2p^{3}\lambda-(1+2p)\lambda^{2}+\lambda^{3}.\label{eq:lambdaHN5}
\end{equation}
Again, as in Eq.~\eqref{eq:PsiLambda}, it is $\Psi\left(p\right)=\log_{2}\lambda$,
which is shown in Fig.~\ref{fig:Plot-of-PsiHN5}. It is remarkable
that, although $\Psi\left(p\right)$ varies smoothly between 0 and
1, near $p=1$ we find only a cubic correction near $p_{c}=1$:
\begin{equation}
\Psi\left(p\right)\sim1-\frac{2}{\ln2}\left(1-p\right)^{3},\qquad p\nearrow p_{c}=1.\label{eq:Psi1Hn5}
\end{equation}

\section{Conclusions\label{sec:Conclusions}}

Our investigation of properties of the cluster formation near the
discontinuous percolation transition in hyperbolic networks affirms
the robustness of the observed finite-size scaling of the largest
cluster in the system. Our study considers more complicated classes
of networks than before, and extends the analysis to include both,
bond- and site-percolation. To obtain our results, we present an automated
means of graph counting, which are essential to accomplish the RG-recursions
for entire functions that are the generators for the cluster sizes.
In the Appendix, we present these methods in somewhat more detail
so that they can serve as a blueprint for similar efforts in the future. 

Our RG study can merely implicate interesting scaling features in
the evolution of the emergent cluster; only detailed simulation can
provide sufficient insight into the mechanics of their formation.
In a parallel effort, we are currently studying bond percolation on
these hyperbolic networks as the familiar limit $q\to1$ of the $q$-state
Potts model \citep{SinghPotts}. In this form, we also hope to better understand the connection
between discontinuous percolation transitions and the phenomenology
of critical transitions as found, for instance, in ferromagnets on
these networks~\citep{BoBr12}, which should be revealed by the interpolation
between $1\leq q\leq2$ in the analytic continuation of the Potts
model.

\section{Acknowledgments\label{sec:Acknowledements}}

We like to thank Trent Brunson, Tomoaki Nogawa, and Takehisa Hasegawa
for fruitful discussions. This work was supported by Grants No. DMR-1207431 
and No. IOS-1208126 from the NSF, and by Grant No. 220020321 
from McDonnel Foundation.

\section{Appendix\label{Appendix}}

\subsection{Automated Graph Counting\label{sub:Automated-Graph-Counting}}

The recursion relations \eqref{eq:MK1rec} for MK1 are obtained by
a process of graph counting depicted in Fig.~\eqref{MK1_3point}.
As the number of possible graphlets increases exponentially for more
complicated hierarchical networks (e.g. HN5 and HNNP), automating
the graph enumeration process $insilico$ makes it easier to obtain
their recursion equations. Key to this process is the adjacency matrix
${\rm {A_{ij}}}$, which gives the information about the presence
of single bonds between two sites in a graph.

\subsubsection{Counting MK1 graphlets:\label{sub:Counting-MK1-graphlets:}}

In the MK1-graphlet in Fig.~\ref{MK1_3point}a, 
\begin{align}
A_{a}=\begin{bmatrix}0 & 1 & 1\\
1 & 0 & 1\\
1 & 1 & 0
\end{bmatrix}\label{Aa}
\end{align}
is an example of an adjacency matrix when all possible bonds are present.
The bonds are bi-directional, which results in a symmetric matrix,
and the diagonal elements are zero, since there are no bonds that
loop back to a site. In the case where two ends are not connected
by a single bond, the adjacency matrix effectively searches for alternate
paths to connect the two end-sites. In Fig~\ref{MK1_3point}e, for
example, the small-world bond is missing, and sites 1 and 3 are not
connected via a single bond. The adjacency matrix is thus,
\begin{align}
A_{e}=\begin{bmatrix}0 & 1 & 0\\
1 & 0 & 1\\
0 & 1 & 0
\end{bmatrix}.
\end{align}

By itself, the adjacency matrix gives the number of one-step end-site
connections. To find the number of two-step end-site connections for
a graphlet, the adjacency matrix must be squared. The off-diagonal
elements of ${\rm {A}^{2}}$ give the number of possible paths between
two sites that are exactly two hops long. Squaring the adjacency matrix
in Fig.~\ref{MK1_3point}a (Eq.~\ref{Aa}) gives
\begin{align}
A_{e}^{2}=\begin{bmatrix}1 & 0 & 1\\
0 & 2 & 0\\
1 & 0 & 1
\end{bmatrix}.
\end{align}
Since matrix element $A_{e,13}^{2}=1$, there exists only one possible
path in which two-steps can be made to connect the end-sites. Since
the maximum path length for the simple case of MK1 is two, only ${ {A_{e,13}}}$
(one step) and $ {A_{e,13}^{2}}$ (two steps) need to be checked
for finding end-to-end connections. 

The graphlets are classified as contributing to $T_{n+1}(x)$ or $S_{n+1}(x,y)$
depending on whether an end-to-end connection exists. The weights
of the graphlets are calculated by first labeling the end-sites as
$x$ and $y$. Both end-sites are labeled $x$ in fully-connected
graphs contributing to $T_{n+1}(x)$, and unconnected graphs contributing
to $S_{n+1}(x,y)$ contain the left end-site labeled $x$ and the
right end-site labeled $y$.

For each graphlet in the $n^{th}$ generation, $x$ or $y$ is assigned
to each site and $T_{n}(x)$ or $S_{n}(x,y)$ to each bond, depending
on whether the end sites are attached. Isolated sites/clusters are
assigned a weight of 1. The contribution of each graphlet in the $(n+1)^{th}$
generation is set as the product of the value assigned to the bonds
and intermediate sites. For example, the two shaded backbone bonds
of Fig.~\ref{MK1_3point}a indicate that the graphlet has two bonds
of type $T_{n}(x)$. The small-world bond exists with probability
$p$, and all the sites are connected to the same cluster. Therefore,
the graphlet contributes to $T_{n+1}(x)$ in the next generation with
weight $p\, x\, T_{n}^{2}(x)$. Similarly, for the graphlet in Fig.~\ref{MK1_3point}f,
the backbone bonds are of the types $T_{n}(x)$ and $S_{n}(x,y)$.
The small-world bond is absent with probability $1-p$, and the end-sites
are connected to separate clusters, $x$ and $y$. Hence, this graphlet
contributes to $S_{n+1}(x,y)$ in the next generation with weight
$(1-p)\, x\, T_{n}(x)S_{n}(x,y)$.

\begin{figure}
\includegraphics[scale=0.5]{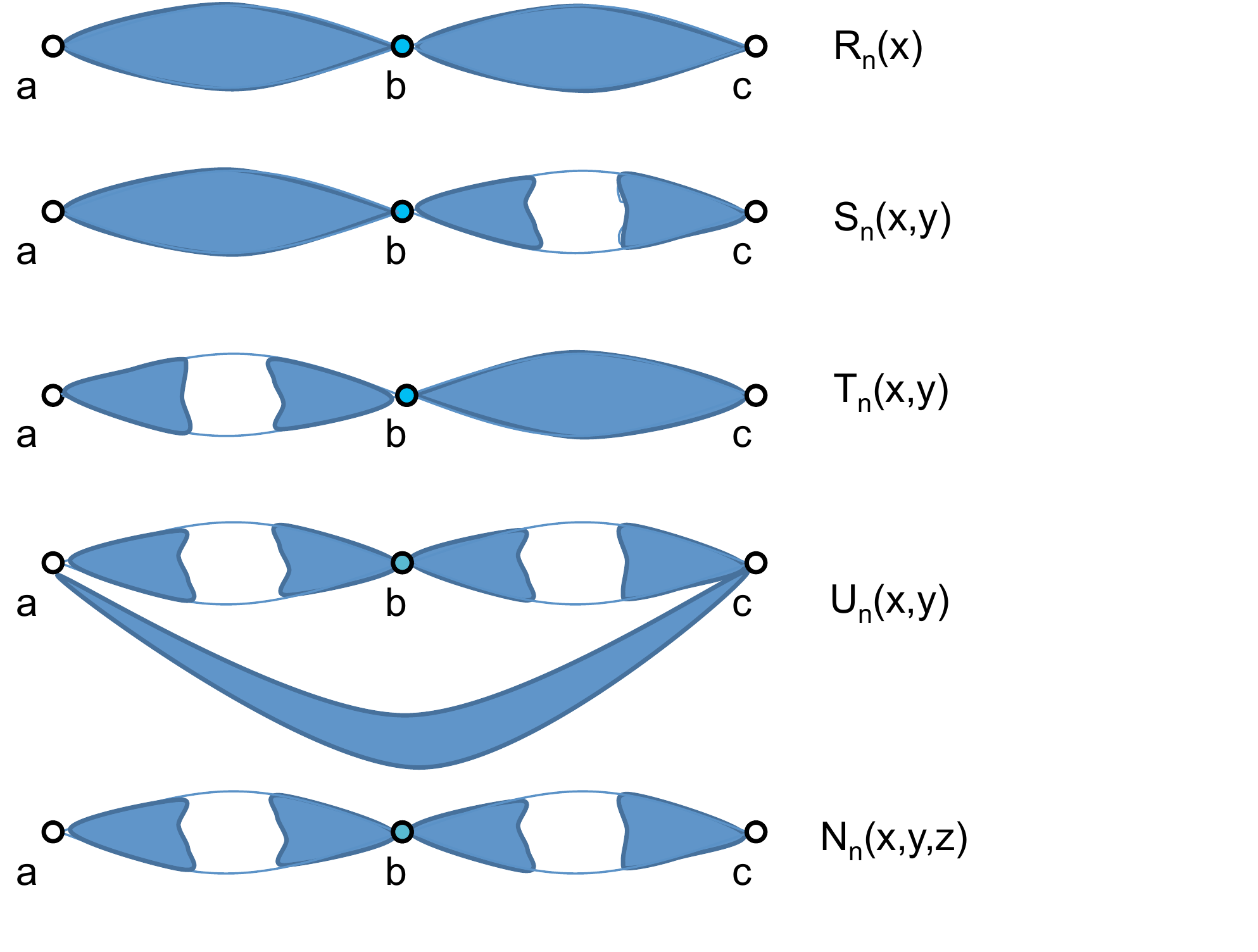} \caption{\label{HNNP_defn} Diagrammatic definition of generating functions
for HNNP and HN5. Sites a, b and c represent the end-sites of the
network. $R_{n}(x)$ consist of one cluster spanning all three end-sites,
$S_{n}(x,y)$, $T_{n}(x,y)$ and $U_{n}(x,y)$ two clusters, one of
which spanning two end-sites, and $N_{n}(x,y,z)$ represents non-spanning
clusters which connect to at most one end-site.}
\end{figure}

\subsubsection{Cluster Generating Function for HNNP:\label{sub:Cluster-Generating-Function HNNP}}

The generating functions for the Hanoi network HNNP in Fig.~\ref{fig:Networks}
can be calculated using the same principles described for MK1. As
in Sec. \ref{sub:Cluster-Generating-function MK1}, we define the
generating functions for HNNP depicted in Fig.~\ref{HNNP_defn}:
\begin{align}
R_{n}(x) & =\;\sum_{k=0}^{\infty}r_{k}^{(n)}(p)x^{k},\label{eq:ClustGenHanoi}\\
S_{n}(x,y) & =\;\sum_{k=0}^{\infty}\sum_{l=0}^{\infty}s_{k,l}^{(n)}(p)x^{k}y^{l},\\
U_{n}(x,y) & =\;\sum_{k=0}^{\infty}\sum_{l=0}^{\infty}u_{k,l}^{(n)}(p)x^{k}y^{l},\\
N_{n}(x,y,z) & =\;\sum_{k=0}^{\infty}n_{k,l,m}^{(n)}(p)x^{k}y^{l}z^{m},
\end{align}
where we introduce the probabilities
\begin{itemize}
\item $r_{k}^{n}(p)$ that sites $a$, $b$ and $c$ are all connected within
the same cluster of size $k$;
\item $s_{k,l}^{n}(p)$ that $a$ and $b$ are mutually connected within
a cluster of size $k$, and $c$ is connected to a separate cluster
of size $l$;
\item $t_{k,l}^{n}(p)$ that $a$ is connected to a separate cluster of
size $k$, and $b$ and $c$ are mutually connected within cluster
of size $l$;
\item $u_{k,l}^{n}(p)$ that $a$ and $c$ are mutually connected within
a cluster of size $k$, and $b$ is connected to a separate cluster
of size $l$;
\item $n_{k,l,m}^{n}(p)$ that $a$ is connected to a cluster of size $k$,
$b$ is connected to a cluster of size $l$, and $c$ is connected
to a cluster of size $m$, but all mutually disconnected.
\end{itemize}
The symmetry of $s_{k,l}^{n}$ and $t_{k,l}^{n}$ are included in
the definition of $S_{n}(x,y)$~\citep{Boettcher09c}. As for MK1,
the three end-notes themselves are not counted in the cluster size.

\begin{figure}
\includegraphics[scale=0.4]{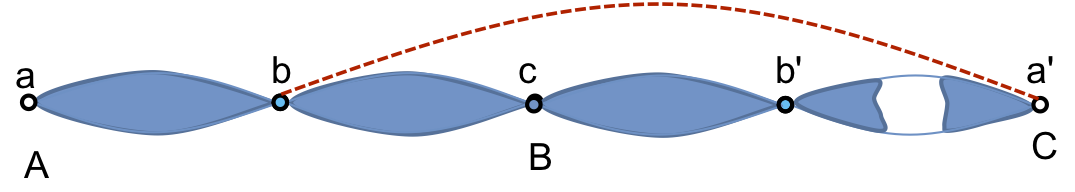} \caption{Example graphlet for HNNP. By looking at the elements of $A_{13}^{2}\;\&\; A_{13}^{2}$
of $A^{2}$, one can see that all the three end-sites are connected.
So this graph contributes to $R_{n+1}(x)$ in the next generation.
In fact all the sites are connected to the same cluster in this case,
which can be verified by looking other element of $A,A^{2},A^{3}$
\& $A^{4}$. Since all sites are connected to the same cluster (say
of size $x$) and there is only one long range small world bond is
present, the weight of the graphlet is $p(1-p)x^{2}R_{n}(x)S_{n}(x,x)/4.$\label{HNNP_Ex1}}
\end{figure}

We want to obtain the system of RG recursions for generating functions,
where $(R_{n+1},S_{n+1},U_{n+1},N_{n+1})$ are functions of $(R_{n},S_{n},U_{n},N_{n};p)$.
The algorithm first generates the adjacency matrices corresponding
to all possible ($2^{8}=256$) graphlets for the HNNP network. For
each one of these graphlets the possibility of their contribution
to one of $(R_{n+1},S_{n+1},U_{n+1},N_{n+1})$ in the next generation
is checked using the adjacency matrices.

As an example of our graph counting algorithm for HNNP, we consider
the graphlet in Fig.~\ref{HNNP_Ex1}. At first glance it appears
that there are two separate clusters of sizes $k$ and $l$. The adjacency
matrix for this graphlet is
\begin{align}
A\ =\ \begin{matrix}{\rm {Node}} & \begin{matrix}a & b & c & b' & a'\end{matrix}\\
\begin{matrix}a\\
b\\
c\\
b'\\
a'
\end{matrix} & \begin{bmatrix}0 & 1 & 0 & 0 & 0\\
1 & 0 & 1 & 0 & 1\\
0 & 1 & 0 & 1 & 0\\
0 & 0 & 1 & 0 & 0\\
0 & 1 & 0 & 0 & 0
\end{bmatrix}
\end{matrix}\label{eq:HNNPAdj}
\end{align}
where the disconnect between sites $a'$ and $b'$ is indicated by
$A_{4,5}=A_{5,4}=0$. After the sites $b$ and $b'$ in Fig.~\ref{HNNP_Ex1}
are decimated in the RG step, the remainder is matched with one of
the graphlets in the generating function diagram in Fig.~\ref{HNNP_defn}.
Thus, only the matrix elements in Eq.~\ref{eq:HNNPAdj} that connect
end sites $a$ to $c$, $a$ to $a'$, and $c$ to $a'$ contribute
to the recursion equations for the generating functions. In general,
the matrix elements for $A^{4}$ must be checked for a five-point
HNNP graphlet, since the maximum number of steps required to connect
all end-sites is four. In our example, 
\begin{align}
A^{4}= & \begin{bmatrix}3 & 0 & 4 & 0 & 3\\
0 & 10 & 0 & 4 & 0\\
4 & 0 & 6 & 0 & 4\\
0 & 4 & 0 & 2 & 0\\
3 & 0 & 4 & 0 & 3
\end{bmatrix}.
\end{align}
Elements $A_{13}^{4}$, $A_{15}^{4}$, and $A_{53}^{4}$ are non-zero,
indicating that the end sites ($a$, $c$, and $a'$) form a contiguous
cluster, where $a^{\prime}$ becomes connected by way of the small-world
bond. The graphlet therefore renormalizes into an $R$-type bond.
To determine its weight, we note that the sites $a$, $b$, and $c$
are connected via an $R_{n}$-type bond and the sites $c$, $b'$,
and $a'$ form an $S_{n}$-type bond. Only the right-hand one of the
small-world bonds is present. Hence, the total weight of this graphlet
in the next generation is $p(1-p)\; x^{2}\; R_{n}(x)S_{n}(x,x)/4$.
Here, $S_{n}$ becomes a function of $x$ in both arguments, since
the small-world bond merges the previously disconnected clusters $x$
and $y$. The factor $1/4$ is due to the symmetry explained in Ref.~\citep{Boettcher09c}. 

This process is repeated for all 256 graphlets with our automated
counting algorithm, where each graphlet is attributed to its appropriate
next-generation graphlet. After adding the weights, the generating
function recursion relations are found to be:%
\footnote{Primed quantities correspond to index $n+1$ and unprimed to $n$.}
\begin{widetext}
\begin{eqnarray}
R^{\prime}(x) & = & \left\{xR\left(x\right)+pxU\left(x,x\right)+\left(1-p\right)U\left(x,1\right)\right\}^{2} +2pxR\left(x\right)\left\{ pxN\left(x,x,x\right)+\left(1-p\right)N\left(x,1,x\right)\right\} \nonumber \\
 &&+pxS\left(x,x\right)\left\{\left(1-p\right)\left[xR\left(x\right)+U\left(x,1\right)\right]+2xR\left(x\right)+pxU\left(x,x\right)\right\} +\frac{3}{4}p^{2}x^{2}S\left(x,x\right)^{2},\label{eq:xyzRGflowHNNP}\\
S^{\prime}(x,y) & = & \frac{1-p}{2}S\left(x,y\right)\left\{px^2S\left(x,x\right)+py^2S\left(y,y\right)+x^2R\left(x\right)+y^2R\left(y\right)+\left(1-p\right)xy\left[R\left(x\right)+R\left(y\right)\right]\right.  \nonumber \\
&&\left.+\left[x+\left(1-p\right)y\right]U\left(x,1\right)+\left[y+\left(1-p\right)x\right]U\left(y,1\right)+p\left[x+y\right]^{2}U\left(x,y\right)+pxN\left(x,1,x\right)+pyN\left(y,1,y\right)\right\} \nonumber \\ 
&&+\frac{p^2}{2}xyS\left(x,y\right)\left\{2U\left(x,y\right)+N\left(x,y,x\right)+N\left(y,x,y\right)\right\}+\left(1-p\right)N\left(x,1,y\right)\left\{p\left[x+y\right]U\left(x,y\right)\right.\nonumber\\
&&\left.+\left(1-p\right)\left[xR\left(x\right)+yR\left(y\right)+U\left(x,1\right)+U\left(y,1\right)\right]\right\}+pxN\left(x,x,y\right)\left\{\left(1-p\right)\left[xR\left(x\right)+U\left(x,1\right)\right]+pyU\left(x,y\right)\right\}\nonumber \\
&&+pyN\left(x,y,y\right)\left\{\left(1-p\right)\left[yR\left(y\right)+U\left(y,1\right)\right]+pxU\left(x,y\right)\right\},\nonumber\\
U^{\prime}(x,y) & = & \frac{1}{4}px\left[\left(2-p\right)x+2\left(1-p\right)y\right]S(x,y)^{2} +pxS(x,y)^{2}\left\{ \left(1-p\right)N\left(x,1,y\right)+pxN\left(x,x,y\right)\right\} ,\nonumber \\
N^{\prime}\left(x,y,z\right) & = & \frac{1}{4}\left(1-p\right)^{2}\left[x+y\right]\left[y+z\right]S\left(x,y\right)S\left(y,z\right) +\frac{1-p}{2}\left[x+y\right]S\left(x,y\right)\left\{ \left(1-p\right)N\left(x,1,z\right)+pxN\left(y,x,z\right)\right\} \nonumber \\
 &  & +\left\{ \left(1-p\right)N\left(x,1,y\right)+pzN\left(x,z,y\right)\right\} \left\{ \left(1-p\right)N\left(x,1,z\right)+pxN\left(y,x,z\right)\right\}  \nonumber \\
 &  & +\frac{1-p}{2}\left[y+z\right]S\left(y,z\right)\left\{ \left(1-p\right)N\left(x,1,y\right)+pzN\left(x,z,y\right)\right\} \nonumber
\end{eqnarray}
Note that for $x=y=z=1$, i.e., when graphlets are counted irrespective
of cluster sizes, these equations revert back to those previously
listed in Ref.~\citep{Boettcher09c}.
\subsubsection{Cluster Generating Function for HN5:\label{sub:Cluster-Generating-FunctionHN5}}
The discussion on how to obtain the RG recursion equations for the
cluster generating functions of HN5 parallels that for HNNP above.
The definition of the generating functions in Eqs.~\ref{eq:ClustGenHanoi},
as illustrated in Fig.~\ref{HNNP_defn}, equally apply to HN5. The
main difference originates with the structure of small-world bonds,
which leads to a planar graph for HN5 and a non-planar graph for HNNP.
Then, our graph counting algorithm results in the following RG recursions:
\begin{eqnarray}
R^{\prime}(x) &=& \left\{U(x,1)+x R(x)\right\}^2 +\frac{1}{2} p^2 x^2 S(x,x)^2 + 2p\left\{N(x,1,x)U(x,1)+xS(x,x)\left[(1-p) U(x,1)+p x U(x,x)\right]\right\} \nonumber \\
&&+pxR(x)\left\{2(1-p)N(x,1,x)+2pxN(x,x,x)+(3-p)xS(x,x)-2U(x,1)+2xU(x,x)\right\}\label{eq:xyzRGflowHN5}\\
S^{\prime}(x,y)&=&(1-p)N(x,1,y)\left\{U(x,1)+U(y,1)+(1-p)\left[xR(x)+yR(y)\right]\right\} \nonumber\\
&&+p(1-p)\left\{x^2R(x)N(x,x,y)+y^2R(y)N(x,y,y)\right\}+\frac{1-p}{4}S(x,y)\left\{px^2S(x,x)+py^2S(y,y)\right\}\nonumber\\
&&+\frac{p(1-p)}{2}\left\{x^2\left[U(x,y)+U(x,x)\right]+y^2\left[U(x,y)+U(y,y)\right]\right\}+\frac{(1-p)^2}{2} \left[x+y\right]\left\{U(x,1)+U(y,1)\right\} \nonumber \\ &&+\frac{1-p}{2}\left\{xR(x)\left[-py+x+y\right]+yR(y)\left[-px+x+y\right]\right\} \nonumber\\
U^{\prime}(x,y)&=&p\left\{N(x,1,y)+\frac{1}{2}(1-p)\left[x+y\right]S(x,y)\right\}^2 +p^2S(x,y)\left\{x^2 N(x,x,y)+y^2 N(x,y,y)\right\} \nonumber\\
&&+\frac{p}{4} S(x,y)^2\left\{\left(1+p-p^2\right)x^2+2p(1-p)xy+(2-p)py^2\right\}\nonumber\\
N^{\prime}(x,y,z)&=&\frac{p(1-p)}{2}\left\{S(x,y)\left[x^2N(y,x,z)+y^2N(y,y,z)\right]+S(y,z)\left[y^2N(x,y,y)+z^2N(x,z,y)\right]\right\}\nonumber\\
&&+\frac{(1-p)^2}{2}\left\{\left[x+y\right]N(y,1,z)S(x,y)+\left[y+z\right]N(x,1,y)S(y,z)\right\}\nonumber\\
&&+\frac{(1-p)}{4}S(x,y)S(y,z)\left\{(1-p)\left[xy+xz+yz\right]+y^2\right\} +(1-p) N(x,1,y) N(y,1,z)\nonumber
\end{eqnarray}
Again, these equations revert back to those previously listed in Ref.
\citep{Boettcher09c} for $x=y=z=1$.
\end{widetext}

\bibliography{/Users/stb/Boettcher} 
\end{document}